\documentclass[prd,letterpaper,onecolumn,preprintnumbers,nofootinbib,superscriptaddress]{revtex4-2}

\usepackage{slashed}
\usepackage{amsmath,amssymb}
\usepackage{graphicx}
\usepackage{units}
\usepackage{bbold}
\usepackage{xcolor}
\usepackage{dsfont}
\usepackage{multirow}
\usepackage[hyperfootnotes=false,colorlinks,citecolor=blue]{hyperref}

\usepackage{comment}
\usepackage[normalem]{ulem}     
\usepackage{physics}
\usepackage{tikz}
\usepackage[compat=1.0.0]{tikz-feynman}

\newcommand{\beq}{\begin{equation}}
\newcommand{\eeq}{\end{equation}}
\newcommand{\bea}{\begin{eqnarray}}
\newcommand{\ena}{\end{eqnarray}}

\def \epsilon {\varepsilon}

\begin{document}
\preprint{CETUP-2024-007, FERMILAB-PUB-24-0559-T, MI-HET-839}
\title{Mass Reconstruction of Heavy Neutral Leptons from Stopped Mesons}
\author{Gustavo F.~S.~Alves}
\affiliation{Instituto de F\'{ı}sica, Universidade de S\~{a}o Paulo, C.P. 66.318, 05315-970 S\~{a}o Paulo, Brazil}
\affiliation{Theoretical Physics Department, Fermilab, P.O. Box 500, Batavia, IL 60510, USA}
\author{P.~S.~Bhupal Dev}
\affiliation{Department of Physics and McDonnell Center for the Space Sciences, Washington University, St.~Louis, Missouri 63130, USA}
\author{Kevin J.~Kelly} 
\affiliation{Department of Physics and Astronomy, Mitchell Institute for Fundamental
Physics and Astronomy, Texas A\&M University, College Station, TX 77843, USA}
\author{Pedro A.~N.~Machado}
\affiliation{Theoretical Physics Department, Fermilab, P.O. Box 500, Batavia, IL 60510, USA}
\begin{abstract}
Heavy neutral leptons (HNLs), depending on their mass and mixing, can be efficiently produced in meson decays from the target or absorber in short- to medium-baseline accelerator neutrino experiments, leaving detectable signals through their decays inside the neutrino detectors. We show that the currently running ICARUS experiment at Fermilab can reconstruct the HNL mass and explore new HNL parameter space in the mass range of 70–190 MeV. The mass reconstruction is enabled by two ingredients: (i) simple two-body kinematics of HNL production from stopped kaon decays at the NuMI absorber, followed by HNL decay into a charged-lepton pair and neutrino at the detector, and (ii) high resolution of Liquid Argon Time Projection Chamber (LArTPC) detectors in reconstructing final state particles. Our mass reconstruction method is robust under realistic energy resolution and angular smearing of the charged leptons, and is applicable to any LArTPC detector. We also discuss the synergy between ICARUS and future facilities like DUNE near detector and PIP-II beam dump in probing the HNL parameter space.
\end{abstract}
\maketitle
\section{Introduction}

The discovery of neutrino oscillations~\cite{Super-Kamiokande:1998kpq, SNO:2002tuh} implies that neutrinos have small but nonzero masses, the origin of which is currently unknown. Nonetheless, this poses a serious challenge to the well-established picture of the Standard Model (SM) of particle physics, whose gauge structure demands exactly massless neutrinos by construction.  Therefore, the explanation for neutrino mass must necessarily come from beyond the SM (BSM) physics. The simplest solution is to extend the SM by adding right-handed neutrinos, thus generating Dirac masses ($m_D$) for the left-handed neutrinos by the Higgs mechanism once electroweak symmetry is broken. However, the extreme smallness of the neutrino mass compared to the charged fermion masses in the SM suggests that another mechanism beyond the standard Higgs mechanism might be at play for neutrino mass generation~\cite{Mohapatra:2006gs}. This is corroborated by the fact that the right-handed neutrinos are SM-gauge-singlets and can therefore acquire Majorana masses ($m_N$), which leads to the seesaw mechanism as a possible explanation of the smallness of neutrino masses: $m_\nu\simeq -m_Dm_N^{-1}m_D^T$~\cite{Minkowski:1977sc, Mohapatra:1979ia, Yanagida:1979as, Gell-Mann:1979vob}. In the minimal version of the seesaw without any additional interactions, the heavy right-handed neutrinos, popularly known as the heavy neutral leptons (HNLs), can only interact with the SM via their mixing with the active neutrinos, which roughly scales as the ratio of the Dirac and Majorana masses: $U\simeq m_Dm_N^{-1}$. Irrespective of the origin of the Majorana mass term and possible ultraviolet completion, even this minimal scenario with just SM+HNLs can not only explain neutrino masses, but also account for the dark matter and baryon asymmetry of the Universe~\cite{Asaka:2005pn}, thus making it a compelling candidate for BSM physics. 

From theoretical considerations, there is no unique preference for the HNL mass scale. Although the original seesaw mechanism assumed the HNL mass scale to be close to the Grand Unification scale, $m_N\sim 10^{14}$ GeV for ${\cal O}(1)$ Dirac Yukawa couplings, it is possible to have $m_N$ much smaller and directly accessible in experiments~\cite{Deppisch:2015qwa,Abdullahi:2022jlv}. In the na\"{\i}ve seesaw limit, the HNL mixing is always suppressed by the light neutrino mass: $|U|^2\simeq m_\nu/m_N$. 
However, this relation is strictly valid only in the single HNL case. With more than one generation of HNLs (and we need at least two to explain the solar and atmospheric mass-squared differences), it is theoretically possible to get arbitrarily large mixing even for small HNL masses by using specific textures of the Dirac and Majorana mass matrices~\cite{Pilaftsis:1991ug,Tommasini:1995ii,Gluza:2002vs,Kersten:2007vk, deGouvea:2007hks, Xing:2009in, Gavela:2009cd, He:2009ua, Adhikari:2010yt, Ibarra:2010xw,  Mitra:2011qr, Lee:2013htl}. The relation between the mixing and the (small) $m_\nu$ can then be avoided by either invoking fine-tuned cancellations or by going to the lepton-number-symmetry-protected limit, which is technically natural. The same symmetry-protected limit can also be achieved by extending the minimal seesaw to e.g. inverse~\cite{Mohapatra:1986aw, Mohapatra:1986bd}, linear~\cite{Wyler:1982dd, Malinsky:2005bi} or extended~\cite{Akhmedov:1995ip, Ma:2009du, Dev:2012sg} seesaw models. 

Without specifying any explicit model constructions, we can treat the HNL mass and mixing as free parameters in a bottom-up phenomenological approach. This is the approach usually taken by the experimental searches of HNLs, which often derive exclusion limits in the HNL mass-mixing plane assuming a single flavor dominance.\footnote{Reinterpreting these limits in realistic neutrino oscillation models often result in weaker constraints, depending on the model details; see e.g., Ref.~\cite{Tastet:2021vwp}.} For a summary of the current HNL limits and future prospects, see, e.g., Refs.~\cite{Atre:2009rg,deGouvea:2015euy,Chrzaszcz:2019inj,Bolton:2019pcu,Abdullahi:2022jlv, Fernandez-Martinez:2023phj} (up-to-date limits are available at~\cite{HNL1, HNL2}). Based on these current limits, the entire experimentally-accessible HNL parameter space can be broadly categorized into 4 mass windows\footnote{For Majorana HNLs coupling to electrons, the neutrinoless double beta could provide the dominant constraint across most of the HNL mass range, but it crucially depends on the Majorana nature of the HNLs and the relative CP phase between the HNLs~\cite{Bolton:2019pcu}.}, namely,
\begin{enumerate}
    \item Low-scale ($\lesssim$ MeV), which is dominated by the cosmological (BBN, CMB) and astrophysical (supernova, $X$-ray) constraints;
    \item Intermediate-scale (${\sim}$MeV -- GeV), driven by experimental searches including meson decays and beam dumps; 
    \item High-scale (${\sim}1$--$100$ GeV), best accessible in searches at high-energy colliders;
    \item Very-high-scale ($\gtrsim$~100 GeV), best tested using precision measurements of electroweak observables.
\end{enumerate}
In this work, we specifically focus on the intermediate mass window of HNLs dominantly coupling to muon neutrinos and show that the accelerator neutrino experiments at the intensity frontier can give world-leading constraints in this mass region.   

Modern accelerator-based neutrino experiments with intense proton beams hitting a fixed target,~e.g.,~with ${\cal O}(10^{21})$ protons-on-target (POT), produce an enormous number of light mesons, particularly pions and kaons, whose subsequent decays into charged leptons and neutrinos can efficiently source the production of HNLs via their mixing with either electron or muon neutrinos. For HNLs lighter than muon mass, muon decays can be an additional HNL source. Once produced, the HNL can travel downstream and decay into visible SM particles leaving detectable signals inside near detectors, typically placed at ${\cal O}(100~{\rm m})$ away from the beam target. 
The accelerator-neutrino probes of HNLs can be further divided into two categories, depending on whether the HNLs are produced from meson decay-in-flight (DIF) or decay-at-rest (DAR). 
In the mass range of interest, namely 20--200~MeV, the DIF experiments include PS191~\cite{Bernardi:1985ny, Bernardi:1987ek},  T2K~\cite{T2K:2019jwa, Arguelles:2021dqn} and MicroBooNE~\cite{MicroBooNE:2023icy}, with MicroBooNE currently setting the best muon-flavored HNL limit for HNL masses above 90 MeV up to 180~MeV.\footnote{Note that the current best limit in the range of 25--90 MeV comes from a phenomenological analysis of the LSND data~\cite{Ema:2023buz}.} The DAR experiments include SIN~\cite{Daum:1987bg}, PIENU~\cite{PIENU:2019usb}, KEK-E89/E104~\cite{Hayano:1982wu, Yamazaki:1984sj}, E949~\cite{E949:2014gsn}, NA62~\cite{NA62:2021bji}, LSND~\cite{Ema:2023buz}, and  MicroBooNE~\cite{Kelly:2021xbv, MicroBooNE:2022ctm}, with LSND~\cite{Ema:2023buz} and E949~\cite{E949:2014gsn} setting the current best limits below 90 MeV and above 180 MeV, respectively.

In this paper, we demonstrate how the two-body kaon decay-at-rest ($K$DAR) channel $K^+\to \mu^+ N$ enables excellent reconstruction of the HNL mass.\footnote{We can also use stopped pion decay $\pi^+\to \mu^+ N$; however, it can only probe a small range of HNL masses $m_N<m_\pi-m_\mu\simeq 30$ MeV, which is already tightly constrained by peak searches in pion decay from PIENU experiment~\cite{PIENU:2019usb}.} The mass reconstruction is enabled due to (i) the two-body kinematics of stopped kaon decays, and (ii) high resolution of LArTPC detectors in reconstructing visible particles. We specifically focus on kaons stopped at the NuMI absorber along the NuMI beamline, which could produce HNLs detectable by the ICARUS experiment at Fermilab, which is one of the largest LArTPC detectors currently operational. Using a few benchmark points, we show that even after accounting for angular-reconstruction effects for the final-state $e^+$ and $e^-$, as well as reconstruction of their energies within the experimental resolution uncertainties, the HNL mass can still be reconstructed with reasonable accuracy from its decay products at the ICARUS detector. This presents a unique opportunity for LArTPC-based neutrino experiments to not only detect HNLs, but also to \emph{measure their masses with high accuracy, even with a handful of HNL events}.

The ICARUS-T600 detector is the largest and farthest detector in the short-baseline neutrino (SBN) program at Fermilab~\cite{ICARUS:2023gpo}. It sits at the confluence of the BNB and NuMI beamlines. For the BNB, ICARUS is an on-axis detector, whereas it is off-axis for the NuMI beam, situated $5.5^\circ$ off-axis and 803~m from the NuMI graphite target, where charged kaons are produced from the 120~GeV protons striking the target. The kaons then travel through a 675~m long decay pipe, with most of the kaons that do not decay in flight being stopped at the absorber located at end of the decay pipe, where they decay at rest.

The NuMI absorber is almost symmetrically located about 100~m from both MicroBooNE and ICARUS detectors. However, ICARUS is approximately five times larger and is expected to have four times more exposure to the NuMI beam than the current MicroBooNE analysis is Ref.~\cite{MicroBooNE:2023icy}. Therefore, we anticipate at least an order of magnitude stronger sensitivity to HNLs in ICARUS compared to MicroBooNE. We evaluate the ICARUS sensitivity to HNLs from $K$DAR  by carefully considering event kinematics and signal efficiencies and find that our results are consistent with this scaling estimate. Specifically, we show that ICARUS will be able to probe previously unexplored HNL parameter space for the muon flavor mixing case in the mass range of 70--190~MeV. In the BNB line, ICARUS is exposed to the 8~GeV BNB beam line located 600~m downstream from the Be target. We also assess ICARUS's sensitivity using kaon decays-in-flight ($K$DIF) produced in the BNB line. However, we find that the reach of NuMI $K$DAR surpasses the sensitivity using the BNB $K$DIF source.

In the event of a positive HNL candidate detection at ICARUS, and after outlining the mass reconstruction measurement and evaluating ICARUS's reach, we next examine how different experiments can complement and confirm the findings from ICARUS. We consider a benchmark scenario for mass reconstruction within the reach of the proposed PIP-II beam dump (PIP2-BD) facility to forecast the number of events it should measure to support ICARUS observation. PIP-II~\cite{Pellico:2022dju} is a major upcoming upgrade to Fermilab accelerator complex, designed to meet the beam requirements of the flagship DUNE experiment. The proposed PIP2-BD~\cite{Toups:2022yxs} aims to probe HNLs similarly to LSND~\cite{Ema:2023buz}. 
With an 800~MeV proton beam and an exposure of $1.2\times 10^{23}$~POT, PIP2-BD can produce a significant number of HNLs from stopped pions and muons at the target. There exists some open HNL parameter space, around 100~MeV, with overlapping ICARUS and PIP2-BD sensitivities. Therefore, should an HNL signal be found in this region by ICARUS, PIP2-BD would be able to confirm it with much higher statistics. Similarly, the future DUNE near detector~\cite{DUNE:2021tad}, will provide further tests of the HNL hypothesis~\cite{Krasnov:2019kdc, Ballett:2019bgd, Berryman:2019dme, Coloma:2020lgy, Breitbach:2021gvv} in the same parameter space explored by ICARUS.

The rest of the paper is organized as follows: In Sec.~\ref{sec:HNL_prod_decay} we discuss the HNL production from stopped kaon decay and the HNL decay to neutrino and charged-lepton pair. 
In Sec.~\ref{sec:mass_recon}, we describe the HNL mass reconstruction, including realistic angular and energy smearing effects. 
In Sec.~\ref{sec:flux}, we present the calculation for the HNL flux and event number for both $K$DAR and $K$DIF sources. 
In Sec.~\ref{sec:icarus}, we present the calculation for deriving the HNL sensitivity at ICARUS, both with the NuMI and BNB beams. 
In Sec.~\ref{sec:pip}, we discuss the synergy between the ICARUS and PIP2-BD projections. Our conclusions are given in Sec.~\ref{sec:con}.

\section{HNL Production and Decay} \label{sec:HNL_prod_decay}

The search for HNLs requires the introduction of a portal through which they communicate with the SM. In this study, we focus on the minimal scenario where HNLs interact with the SM through mixing with the SM neutrinos. Specifically, we consider one additional Majorana HNL that primarily mixes with muon neutrinos:
\begin{equation}
    \ket{\nu_\mu} = \sum_i^3 U^*_{\mu i}\ket{\nu_i} + U_{\mu 4}^* \ket{N} , 
\end{equation}
where $U_{\mu i}$ with $i=1,2,3$ are the muon-row elements of the leptonic mixing matrix parameterized in terms of the 3-neutrino oscillation parameters~\cite{nufit} and $U_{\mu 4}$ is the unknown mixing with the HNL. For simplicity, we assume that any other HNLs, necessary to properly generate the light neutrino parameters, are either outside the mass range of interest or do not primarily couple to the muon flavor, and thus do not affect our analysis.
\begin{figure}[t!]
    \centering
    \begin{tikzpicture}
      \begin{feynman}
        \vertex (a1) {\( \)};
               \vertex [below right =of a1] (a2);
        \vertex [right=of a2] (a3);
        \vertex [above right=of a3] (a4) {\(N\)};
        \vertex [below right=of a3] (a5) {\(\mu\)};
    
        \vertex[below=6em of a1] (b1) {\( \)};
         
        \diagram*{
          (a1) -- [fermion] (a2),
          (a2) -- [boson, edge label'=\(W\)] (a3),
          (a3) -- [fermion] (a4),
          (a3) -- [anti fermion] (a5),
          (b1) -- [anti fermion] (a2),
        };

       \draw[fill=black] (0,-1) ellipse (.2cm and 1cm) node [pos=10, left] {\(M\)};
      \end{feynman}
    \end{tikzpicture}
    \hspace{0.5cm}
     \begin{tikzpicture}
        \begin{feynman}
            \vertex (a) {\(N\)};
            \vertex [right=of a] (b);
            \vertex [above right=of b] (f1) {\(\nu_{\mu}\)};
            \vertex [below right=of b] (c);
            \vertex [above right=of c] (f2) {\(e^+\)};
            \vertex [below right=of c] (f3) {\(e^{-}\)};
            \diagram* {
                (a) -- [fermion] (b) -- [fermion] (f1),
                (b) -- [boson, edge label'=\(Z\)] (c),
                (c) -- [anti fermion] (f2),
                (c) -- [fermion] (f3),
                };
        \end{feynman}
    \end{tikzpicture}
    \caption{Feynman diagrams for production of muon-flavored HNL from charged meson decay (left) and the decay of HNL into $e^+e^-\nu_\mu$ (right) via neutral-current. For heavier HNLs, additional decay modes $N\to \mu^+\ell^-\nu_\ell$ (with $\ell=e,\mu$) open up via charged-current, not shown here.}
    \label{fig:HNL_production}
\end{figure}
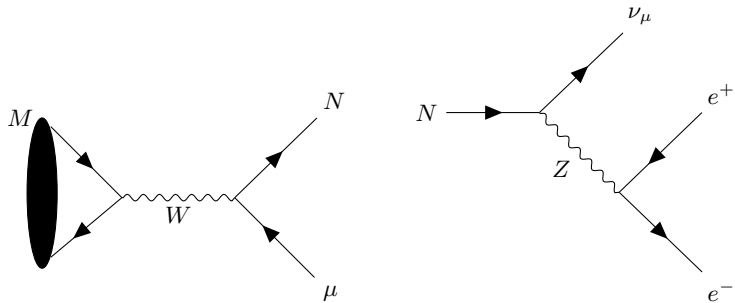

The mixture is a well-studied mechanism for producing and detecting HNLs.  If kinematically possible, HNLs can be produced from meson decays, as shown in the left panel of Fig.~\ref{fig:HNL_production}. The branching ratio (BR) for this process~\cite{Shrock:1980vy}  is given by
\begin{equation}
    \text{BR}(M \to \mu N) =  |U_{\mu 4}|^2 \rho_N(x_\mu,x_N)\text{BR}(M \to \mu \nu_{\mu}),
\end{equation}
where $x_N = (m_N/m_M)^2$, $x_\mu = (m_\mu/m_M)^2$ and $\rho_N(x,y) = [x_\mu + x_N - (x_\mu - x_N)^2]\lambda^{1/2}(1, x_\mu, x_N)/[x_\mu(1- x_\mu)^2]$ is a phase space factor, with $\lambda(a,b,c) = a^2 + b^2 + c^2 - 2(ab + bc + ac)$. In the limit $m_{N}\ll m_M$, we have $\rho_N\to 1$. In the following, we will mainly focus on the kaon contribution, i.e., $M=K$. Below the muon threshold, HNLs can also be produced from muon decays resulting from meson decays $\mu \to N e \nu_e$. Although we will comment on the contributions from pions and muons at the end, the kaon contribution is the dominant source in the mass range of interest.

For detection, depending on $m_N$, several decay channels are available: (i) $N \to 3\nu$, (ii) $N\to \nu \gamma$, (iii) $N\to \nu \pi^0$, (iv) $N\to \mu^\pm \pi^\mp$, (v) $N\to \nu \mu^\pm e^\mp $, and (vi) $N\to \nu \mu^+\mu^-$. Channel (i) is completely invisible and thus not detectable. Channels (ii) and (iii) also present challenges due to the large $\pi^0\to \gamma\gamma$ background. Channels (iii)-(vi) require the HNLs to be above the muon or pion threshold. A complete list of the decay modes can be found in Refs.~\cite{Gorbunov:2007ak, Asaka:2012bb}. 

We are primarily interested in HNLs with masses in the 20--200~MeV range. Within this range, the decay $N\to \nu e^+e^-$ is the best visible candidate, despite having a branching ratio of about 10\%. The remaining ${\sim}90\%$ of decays are mostly $N\to 3\nu$ for $m_N<m_{\pi}$ and $N\to \nu \pi^0$ for $m_N>m_{\pi}$.  The decay $N \to \nu_\mu e^- e^+$, proceeds exclusively through neutral currents if the HNL mixes only with the muon, as illustrated in the right panel of Fig.~\ref{fig:HNL_production}. The partial decay rate into this channel, up to corrections proportional to the electron mass, is given by~\cite{Gorbunov:2007ak}
\begin{equation}
    \Gamma(N \to \nu_\mu e^+ e^-) = 2 \frac{G_F^2 |U_{\mu 4}|^2 m_N^5}{768 \pi^3} (1 - 4 s_W^2 + 8 s_W^4)+{\cal O}\left(\frac{m_e^2}{m_N^2}\right) \, ,
\end{equation}
where $s_W$ is the sine of the weak mixing angle and the factor of 2 is present for Majorana HNLs.\footnote{Since we are not relying on the lepton-number-violating nature of the signal, our analysis is equally valid for pseudo-Dirac HNLs as well, modulo the factor of 2 difference, depending on whether either or both of the HNLs in the pseudo-Dirac pair couple dominantly to muon flavor.} 

The $N \to \nu \pi^0$ presents an exciting opportunity when $m_N > m_{\pi^0}$ as the expected branching ratio into this channel is relatively large for a large range of $N$ masses. Given that more analyses in the experimetnal literature have been performed to search for novel $e^+ e^-$ signatures (due to motivations from LSND/MiniBooNE anomalies) than for $\pi^0$ emerging from an HNL decay, we will restrict our focus to the $N \to \nu e^+ e^-$ channel for the present work. Nevertheless, many of the strategies we present, including precise HNL mass reconstruction, are also possible in the $N \to \nu \pi^0$ channel. Since the HNL is very long-lived in our case, including the $N \to \nu \pi^0$ channel will only \textit{increase} our sensitivity.

\section{Reconstructing the HNL mass}
\label{sec:mass_recon}
\begin{figure}[b]
    \centering
    \includegraphics[width = 0.9\textwidth]{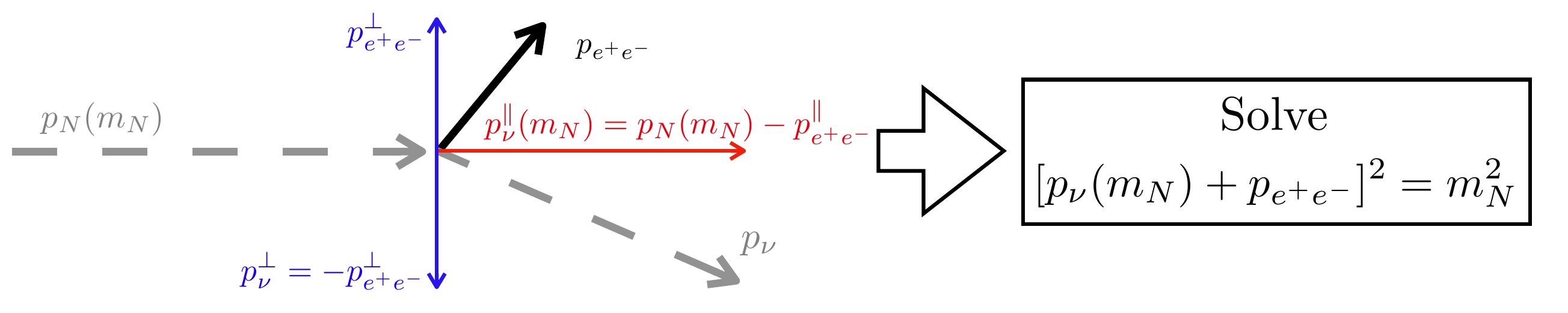}
    \caption{Schematic on how to reconstruct the mass of the HNL.}
    \label{fig:mN_reconstruction}
\end{figure}

One of the advantages of using two-body DAR source to produce the HNLs is that we can unambiguously reconstruct the HNL mass from simple kinematics in the event of a positive HNL signal. For concreteness, let us consider $K$DAR source: $K^+\to \mu^+N$, and the decay channel $N \to \nu e^+ e^-$. Then we can reconstruct the HNL mass by following a few simple steps (see Fig.~\ref{fig:mN_reconstruction}): 
\begin{enumerate}
    \item From the two-body kinematics $K^+ \to \mu^+N$ for a kaon at rest, we know  the four-momentum $p_N$ of the HNL as function of $m_N$, that is, $p_N(m_N)$, given $m_K$ and $m_\mu$. 
    \item Measure the electron-positron pair momentum. This sets the light neutrino transverse momentum $p_{\nu}^\perp=-p_{e^+e^-}^\perp$.
    \item Find the light neutrino momentum parallel to the $N$ direction by using momentum conservation in $N$ decay, i.e., $p_\nu^\parallel(m_N) = p_{N}(m_N)- p^\parallel_{e^+ e^-}$. Note that it will be a function of $m_N$.
    \item Solve $\left[p_\nu(m_N)+p_{e^+e^-}\right]^2=m_N^2$ to find the mass of the HNL.
\end{enumerate}
\begin{figure}[t!]
    \centering
    \includegraphics[width = 0.7\textwidth]{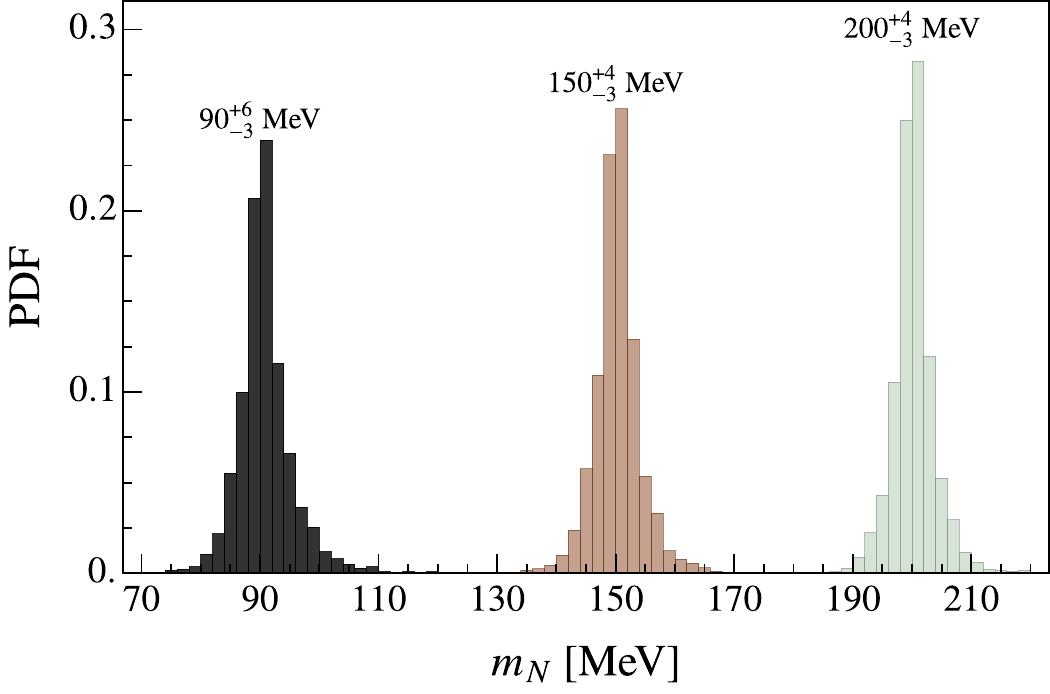}
    \caption{$K$DAR-ICARUS mass reconstruction capability for three benchmark values of the HNL mass $m_N = 90, 150, 200~\rm{MeV}$, including a $3^\circ$ angular smearing in each of the electromagnetic showers.}
    \label{fig:mN_rec}
\end{figure}
\begin{figure}[t!]
    \centering
    \includegraphics[width = 0.7\textwidth]{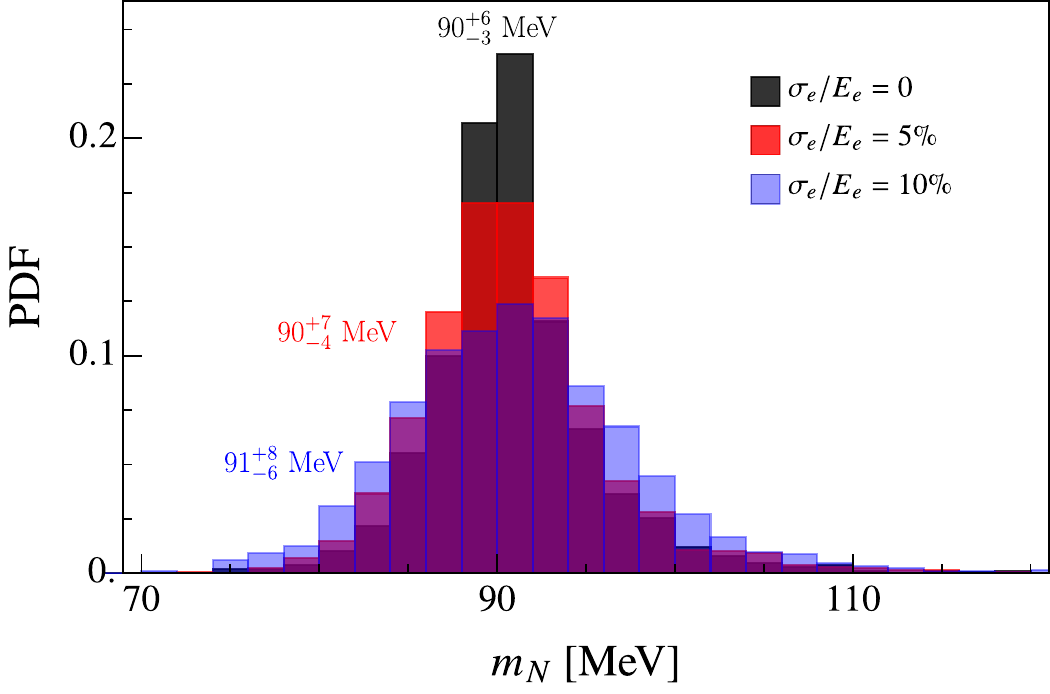}
    \caption{$K$DAR-ICARUS mass reconstruction capability for a benchmark value of the HNL mass $m_N = 90~\rm{MeV}$, including a $3^\circ$ angular smearing in each of the electromagnetic showers, and with 0\%, 5\% and 10\% energy smearing.}
    \label{fig:mN_rec_energy}
\end{figure}
We show in Fig.~\ref{fig:mN_rec} the result of this analysis for three benchmark values of the HNL mass $m_N = 90, 150, 200~\rm{MeV}$. We apply a $3^\circ$ angular uncertainty to each of the electromagnetic showers, following Ref.~\cite{MicroBooNE:2019rgx}. We fitted the probability distribution functions (PDFs) to a Gaussian and found the best-fit values and $1\sigma$ error bars on the masses, as shown in the Figure. It is clear that the reconstructed masses match pretty well with the true values.

We also consider the effect of energy smearing, on top of the $3^\circ$ angular smearing. We compare the PDFs with no energy smearing, and Gaussian smearing, defined by $\exp{-(E_e-E_e^{\rm rec})^2/2\sigma_e^2}/(\sqrt{2\pi}\sigma_e)$, with $\sigma_e/E_e= 5\%$ and $10\%$, following Refs.~\cite{Friedland:2018vry, Kelly:2021jfs}. 
Here $E_e$ and $E_{e}^{\rm rec}$ are the true and  reconstructed values of the electron/positron energy.
The corresponding reconstructed HNL mass values for the 3 benchmarks shown in Fig.~\ref{fig:mN_rec} are reported in Table~\ref{tab:mass}.  
In Fig.~\ref{fig:mN_rec_energy}, we show the corresponding PDFs for a benchmark value of $m_N=90$ MeV. When performing all of these simulations, the above mentioned reconstruction effects (on the direction and energy) are performed on the final-state electron and positron \textit{independently}, and then the reconstructed $p_{e^+ e^-}$ is determined from those (smeared) objects.
This illustrates the point that the HNL mass reconstruction at $K$DAR-ICARUS is robust against both angular and energy smearing effects. Also, note that the mass reconstruction can be done on an event-by-event basis, so it works even in a low-statistics scenario. 
\begin{table}[h!]
\centering
    \begin{tabular}{|c|c|c|c|}
    \hline
        True Mass [MeV]& Reconstructed Mass [MeV]  & Reconstructed Mass [MeV] & Reconstructed Mass [MeV] \\ 
          &  $\sigma_e/E_e= 0$ &  $\sigma_e/E_e = 5\%$ &  $\sigma_e/E_e = 10\%$ \\ \hline
        90 & $90^{+6}_{-3}$ & $90^{+7}_{-4}$ & $91^{+8}_{-6}$ \\ \hline
        150 & $150^{+4}_{-3}$ & $151^{6}_{-5}$ & $151^{+10}_{-8}$ \\ \hline
        200 & $200^{+4}_{-3}$ & $201^{+7}_{-6}$ & $201^{+10}_{-10}$ \\ \hline
    \end{tabular}
\caption{True values and their corresponding reconstructed values of the HNL mass with different Gaussian energy smearing.}
\label{tab:mass}
\end{table}

Throughout this section, we have assumed that only signal events are analyzed. Of course, background events, mostly coming from misidentifying one or two photons from a NC$\pi^0$ event as a two-electron shower, can contaminate such a search. Nevertheless, there are two handles that can be used to mitigate these backgrounds. First, the two electromagnetic showers should reconstruct the $\pi^0$ mass. Second, the aforementioned procedure to reconstruct the $N$ mass, when applied to NC$\pi^0$ events, should mostly result in a flat spectrum, i.e.~not exhibiting any significant $m_N$ peak. Regardless, we encourage future studies on the reconstruction capabilities of LArTPC detectors in these event topologies.

\section{HNL Flux and Event Rate}
\label{sec:flux}

To assess the reach of an experiment in the search for HNLs, three key components are required: the HNL flux, its decay channels, and the detector's ability to detect particles produced by HNL decays. In this section, we discuss how to get the HNL flux and present the primary formula for calculating the HNL event rate in accelerator neutrino experiments. We follow closely the discussion from Ref.~\cite{Asaka:2012bb}, which offers an analytical approach to deriving the HNL flux for T2K, considering $K$DIF. We focus on neutrinos produced in two-body meson decays, which constitute the majority of the neutrino beam in accelerator experiments, though the formalism can be easily adapted to other scenarios.

Our first goal is to determine the flux of daughter neutrinos. For now, we consider SM neutrinos and later discuss how to adapt the formalism for HNLs. We begin by assuming that we have access to the number of mesons $M$ produced in the interactions of protons with the target material, per momentum interval $(p_M, p_M + dp_M)$, i.e., $\mathcal{N}(p_M) = dN_M/dp_M$. If we integrate this quantity over the magnitude of the 3-momentum $p_M$ we find the total number of mesons $N_M$ produced in the primary proton collision. As the mesons propagate, $\mathcal{N}(p_M)$ changes due to decay. We define the position-dependent number of mesons per unit energy and propagation distance as
\begin{equation}
         \frac{d \mathcal{N}(p_M,x)}{dx} = \frac{\mathcal{N}(p_M,0)}{\Lambda_M}\, e^{-\frac{x}{\Lambda_M}},
    \label{eq:pos_dependent_mother_flux}
\end{equation}
where $x$ is the propagating distance and $\Lambda_M = p_M/(m_M \Gamma_M)$ is the lab-frame meson decay length, with $\Gamma_M$ being its total width. 

We define the differential branching fraction of a meson with momentum $p_M$ in the lab frame, decaying into a neutrino with energy in the range $(E_\nu, E_\nu + dE_\nu)$, and within the solid angle defined by the intervals $(\theta, \theta + d\theta)$ for the polar angle and $(\phi, \phi + d\phi)$ for the azimuthal angle as
\begin{equation}
   \frac{d\mathcal{B}(E_\nu,\theta,\phi)}{dE_\nu d\cos \theta d \phi} = \frac{1}{\Gamma_{M}}\frac{d^3 \Gamma_{M \to \nu + X}}{dE_\nu d\cos{\theta} d \phi},
\end{equation}
with $\Gamma_{M \to \nu + X}$ being the partial decay width of the two-body meson decay $M \to \nu X$ in the lab frame~\cite{Asaka:2012bb}:
\begin{equation}
    \frac{d^3 \Gamma_{M \to \nu + X}}{dE_\nu d\cos{\theta} d \phi} = \frac{|\mathcal{M}|^2}{2 E_M} \frac{E_\nu}{8\pi^2}\delta(m_M^2 - m_X^2 - 2 E_\nu E_M + 2 p_M E_\nu \cos{\theta}),
    \label{eq:SM_nu_prod}
\end{equation}
where $|\mathcal{M}|^2 = 2G_F^2 f_M^2 m_M^4 |V_{ij}|^2(m_X^2/m_M^2 - m_X^4/m_M^4)$ is the matrix element for two-body decay, and $E_M=\sqrt{p_M^2+m_M^2}$ is the energy of the meson. Here, $f_M$ is the meson decay constant, $G_F$ is the Fermi constant, $V_{ij}$ is the relevant CKM matrix element ($V_{ud}$ for pion decay and $V_{us}$ for kaon decay), and $m_X$ is the mass of the charged lepton produced alongside the SM neutrino, which is a muon in our case. We note that a two-body decay should depend on only two kinematical variables. The delta function is introduced in Eq.~\eqref{eq:SM_nu_prod} to account for the correlation between $E_\nu$ and $\theta$. However, treating the decay width as a function of three variables is useful for calculating the number of events~\cite{Asaka:2012bb} as we comment next. 

For a detector located at a distance $d$ from the target material, and with a decay volume of length $l$ along the direction of propagation, the neutrino flux reaching the detector is given by
\begin{equation}
    \phi_\nu(E_\nu) = \int_0^l dx \int_{-1}^1 d\cos{\theta}\int_0^{2\pi}d\phi \int_{0}^\infty dp_M\frac{1}{4 \pi d^2}\frac{d \mathcal{N}(p_M,x)}{dx} \frac{d\mathcal{B}(E_\nu,\theta,\phi)}{dE_\nu d\cos \theta d \phi}  \Theta(\theta,\phi),
    \label{eq:light_neutrino_flux}
\end{equation}
where $\Theta(\theta, \phi)$ is a Heaviside function, equal to one if the neutrino's kinematic configuration allows it to reach the detector, and zero otherwise. At this stage, keeping the differential branching as a function of three variables proves useful. Details on how to apply the kinematical cuts are thoroughly described in Ref.~\cite{Asaka:2012bb}. The geometric area factor $4\pi d^2$ can be replaced by the detector's effective area to account for efficiencies, such as the dependence on solid angles. However, we will not include this adjustment in our simulation as we do not have access to this information. The flux derived in Eq.~\eqref{eq:light_neutrino_flux} for SM neutrinos can be readily adapted for HNLs by replacing the partial decay width in Eq.~\eqref{eq:SM_nu_prod} with the one relevant for HNLs. The corresponding expressions can also be found in~\cite{Asaka:2012bb}. 

Now, we turn to the discussion of the number of events. The detector measures the decay products of the HNL. We denote the HNL flux by $\phi_N(E_N)$, but our focus is on how this flux evolves as the HNL propagates. For that, we introduce the position-evolved HNL flux 
\begin{equation}
    \phi_N(E_N, x) = \phi_N(E_N,0) e^{-\frac{x}{\Lambda_N}},    
\end{equation}
such that the number of events in the channel $N \to j$ is given by
\begin{equation}
    N_j = A \int_{m_N}^\infty d E_N\int_{x_i}^{x_f} dx\, \phi_N(E_N,x)\frac{\text{BR}(N\to j)}{\Lambda_N},
\end{equation}
where $\Lambda_N$ is the decay length of the HNL, $\text{BR}(N \to j)$ is its branching ratio into the channel $j$, $A$ is the detector’s cross-sectional area, and $x_i$ and $x_f$ are the positions marking the beginning and end of the detector in the direction of the HNL propagation. 

If $\Lambda_N$ significantly exceeds both the distance to the detector entrance $x_i$ and the detector's dimension in the direction of the HNL propagation $L = x_f - x_i$, i.e., $\Lambda_N \gg x_i$ and $\Lambda_N \gg L$, we can approximate $\phi_N(E_N, x) \approx \phi_N(E_N, 0) \equiv \phi_N(E_N)$. In this case, the number of events reduces to~\cite{Asaka:2012bb}
\begin{equation}
    N_j = V\int_{m_N}^\infty dE_N \xi(E_N)\frac{\phi_N(E_N)}{\lambda_j},
    \label{eq:nevts_HNL}
\end{equation}
where $\lambda_j=\Lambda_N/\text{BR}(N \to j)$ is the partial decay length and $V = A L$ is the fiducial detector volume. We also include an energy-dependent efficiency factor, $\xi(E_N)$, to account for the detector's ability to reconstruct particles produced in the HNL decay. Under this approximation, the number of events is independent of the HNL lifetime and we have verified numerically that the parameter space considered in this work falls within the range of validity for this approximation. 

To illustrate how this formalism can be applied to obtain the number of HNL events, we first consider decays at rest. In this scenario, the HNL flux is given by
\begin{align}
    \begin{split}
        \phi_N(E_N) &= \int_{-1}^1 d\cos{\theta}\int_0^{2\pi}d\phi \int_{0}^\infty dp_M\frac{1}{4 \pi d^2}\mathcal{N}(p_M,0)e^{-\frac{l}{\Lambda_M}}\frac{1}{\Gamma_{M}}\frac{d \Gamma_{M \to \nu + X}}{dE_N d\cos{\theta} d\phi}\Theta(\theta,\phi)\\
        &= N^{\text{DAR}}_M\frac{\Delta \theta \Delta  \phi}{4 \pi d^2} \frac{1}{\Gamma_{M}}\frac{d \Gamma_{M \to \nu + X}(m_M)}{dE_N}, 
    \end{split}
\end{align}
where we have utilized the fact that the decay width for a particle at rest is independent of $\theta$ and $\phi$. Consequently, the Heaviside function projects the number of events into the solid angle $\Delta \theta \Delta \phi$ covered by the detector. $N^{\text{DAR}}_M$ is the number of mesons that reach the absorber, located at a distance $l$ from the target:
\begin{equation}
   N^{\text{DAR}}_{M} = \int_{0}^\infty d p_M\mathcal{N}(p_M,0)e^{-\frac{l}{\Lambda_M}},
\end{equation}
such that the number of HNL events from decays at rest in the channel $j$ is
\begin{equation}
    N_j = N^{\text{DAR}}_M \frac{A}{4 \pi d^2} \int_{m_N}^\infty dE_N \frac{L}{\lambda_j(E_N)}\frac{1}{\Gamma_M}\frac{d \Gamma_{M \to \nu + X}(m_M)}{dE_N}.
\end{equation}
We included the solid angle covered by the detector in the definition of its cross sectional area $A$. Usually, the number of events from mesons decay at rest is expressed in terms of the partial decay time $\tau_j$, which is related to the partial decay length by $\lambda_j(E_N) = \gamma \beta \tau_j$, where $\gamma$ and $\beta$ are the Lorentz boost and speed of $N$ in the lab frame, respectively. For a two-body meson decay $M \to \mu N$ at rest, the decay width contains a Dirac $\delta$-function which fixes the HNL energy:
\begin{equation}
    E_N = \frac{m_M^2 + m_N^2 - m_\mu^2}{2m_M},
    \label{eq:EN}
\end{equation}
and the number of events simplifies further to
\begin{equation}
    N_j = N_M \frac{A}{4 \pi d^2} \frac{L}{\gamma \beta \tau_j } \text{BR}(M \to \nu + X),
    \label{eq:master_formula} 
\end{equation}
which recovers the results from Refs.~\cite{Kelly:2021xbv,Ema:2023buz}. 

\begin{figure}[t!]
    \centering
    \includegraphics[width = 0.65\textwidth]{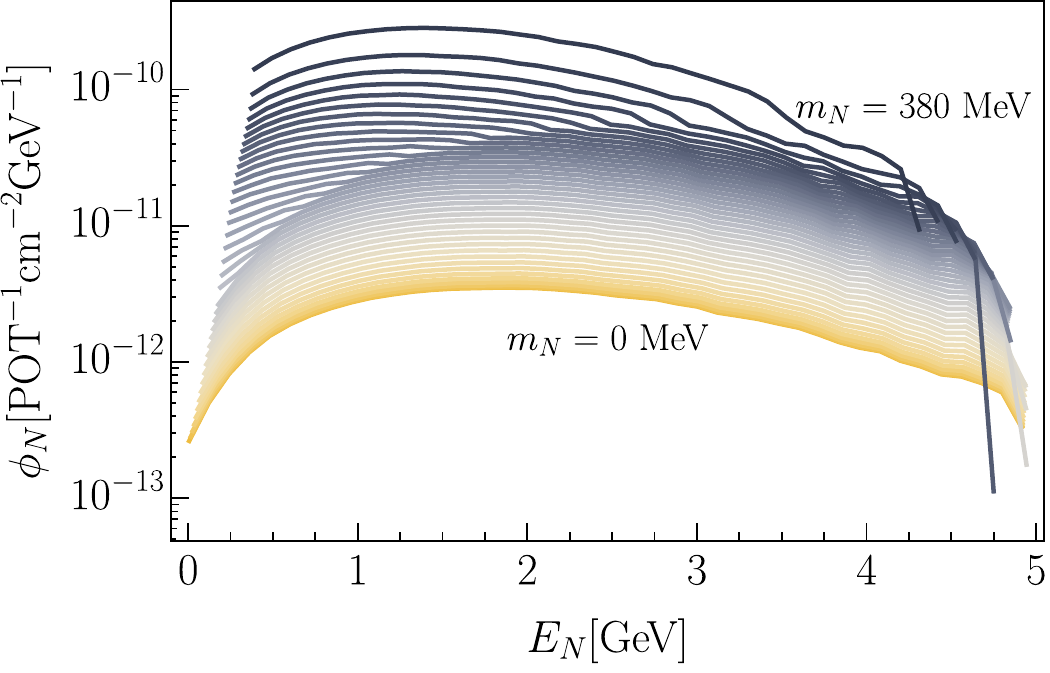}
    \caption{HNL fluxes from $K$DIF for ICARUS considering the BNB line for different values of the HNL mass ($m_N=0$ at the bottom to $m_N=380$ MeV at top, in steps of 10 MeV). The flux is normalized by $|U_{\mu 4}|^4$.}
    \label{fig:flux_HNL}
\end{figure}

To obtain analytically either $N^{\text{DAR}}_M$ or the number of mesons decaying in flight we need to determine the number of mesons produced per momentum interval $\mathcal{N}(p_M)$. In Ref.~\cite{Asaka:2012bb}, this is obtained by fitting Eq.~\eqref{eq:light_neutrino_flux} to the SM neutrino flux from T2K simulations. We do the same but for the BNB-ICARUS setup. The HNL fluxes we obtain, for $|U_{\mu N}|^2 =1$, are shown in Fig.~\ref{fig:flux_HNL}. We focus only on the kaon contribution as our goal is to make a comparison between the reach of $K$DAR-ICARUS for the NuMI line with $K$DIF-ICARUS for the BNB line. Moreover, the kaon contribution dominates in the allowed mass range of interest. We start with the kaon production cross section $d\sigma/dp_K$ for the BNB line. This is obtained by rescaling the measurements of the cross section at higher energies to the BNB line proton beam energy using the Feynman scaling parametrization, as detailed in Ref.~\cite{Mariani:2011zd}. They provide the invariant $K^+$ production cross sections as a function of the kaon momentum for several values of the angle $\vartheta_K$ of the kaon relative to the incident proton beam. For our analysis, we will use the $\vartheta_K = 0.075$~rad result. However, our results are not sensitive to this choice since we will fit the SM neutrino flux using the kaon cross section, and varying $\vartheta_K$ yields similar results. 

From the production cross section, we determine $\mathcal{N}(p_K)$ by including two adjustments: a normalization factor and an offset parameter~\cite{Asaka:2012bb}. The normalization factor accounts for the number of kaons produced and the effective target area. The offset parameter accounts for potential deformations in $\mathcal{N}(p_M)$, such as those caused by the magnetic horns.
We fix both parameters by computing the SM neutrino flux using Eq.~\eqref{eq:light_neutrino_flux} and adjusting the normalization and offset to match the SM neutrino flux from $K$DIF, as estimated for the MiniBooNE experiment~\cite{MiniBooNE:2008hfu}. In possession of $\mathcal{N}(p_K)$, we can determine the HNL fluxes shown in Fig.~\ref{fig:flux_HNL}. For these calculations, we set the HNL mixing to one, meaning that to calculate the number of events, the flux must be scaled by $|U_{\mu N}|^4$.

\section{ICARUS Sensitivity} 
\label{sec:icarus}

ICARUS is the largest LArTPC detector currently operational in a neutrino beam~\cite{ICARUS:2004wqc}. It is collecting data from both the 8~GeV BNB and the 120~GeV NuMI beams at Fermilab~\cite{ICARUS:2023gpo}. The detector consists of two semi-independent, symmetric modules filled with liquid argon. Each module~\cite{MicroBooNE:2015bmn} has an active volume measuring approximately 3.2~m in height, 2.96~m in width, and 18~m in length. The combined front cross-sectional area of the two modules is $A= 2\times 3.2 \times 2.96 = 18.9~{\rm m}^2$, with a total detector volume of 340~$\text{m}^3$. For a view of the experimental setup, see e.g., Refs.~\cite{MicroBooNE:2015bmn, Batell:2019nwo}. 

\begin{figure}[!t]
    \centering
    \includegraphics[width= 0.70\textwidth]{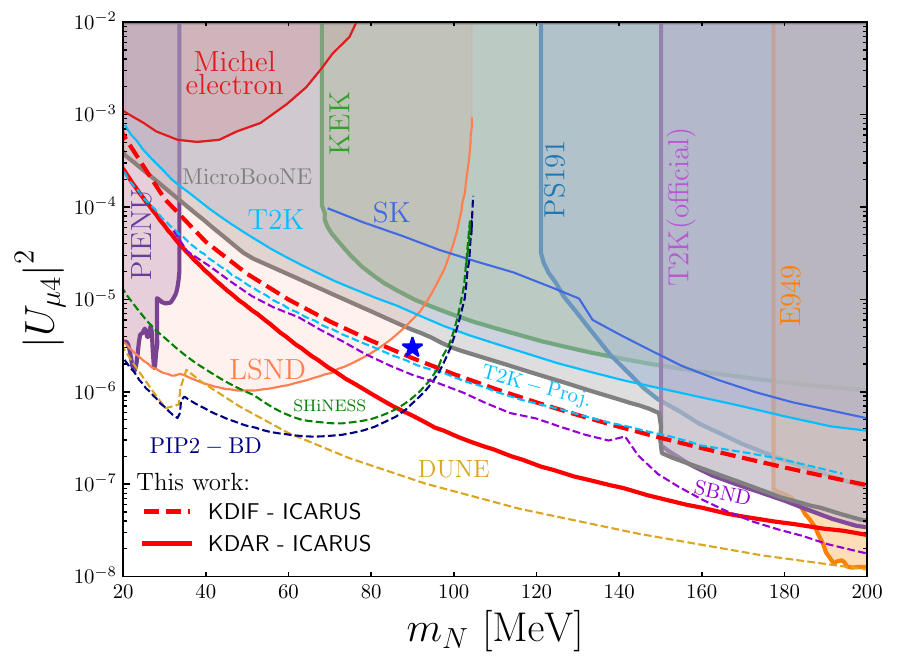}
    \caption{The $K$DAR-ICARUS (red solid) and $K$DIF(BNB)-ICARUS (red dashed) sensitivities at 95\% C.L. in the muon-flavored HNL mass-mixing plane. For comparison, we also show the relevant existing experimental constraints from PIENU~\cite{PIENU:2019usb} (90\% C.L.), KEK~\cite{Hayano:1982wu} (90\% C.L.), PS191~\cite{Bernardi:1987ek} (90\% C.L.),  T2K~\cite{T2K:2019jwa} (90\% C.L.), E949~\cite{E949:2014gsn} (90\% C.L.) and MicroBooNE~\cite{MicroBooNE:2023icy} (90 \% C.L.) by the shaded regions which are ruled out. Other existing constraints (not by experimental collaborations) from Michel electrons in muon decay~\cite{deGouvea:2015euy} (99\% C.L.), LSND~\cite{Ema:2023buz} (90\% C.L.), T2K~\cite{Arguelles:2021dqn} (90\% C.L.) and Super-K~\cite{Coloma:2019htx} (90\% C.L.) are also shown by thin solid curves, which might be ruled out. Future projections from T2K~\cite{Arguelles:2021dqn} (90\% C.L.), SBND~\cite{SBND} (90\% C.L.), DUNE near detector (DUNE-ND)~\cite{Berryman:2019dme} (99\% C.L.), PIP2-BD~\cite{Ema:2023buz} (90\% C.L.) and SHiNESS~\cite{Soleti:2023hlr} (90\% C.L.) are also shown.  We take the point represented by the blue star as a benchmark event in ICARUS to forecast the number of events PIP2-BD would see. 
    }
    \label{fig:summary}
\end{figure}

In this section, we derive the HNL sensitivities for the ICARUS-T600 detector using two different configurations: (i) HNLs produced in $K$DAR from the NuMI absorber, which is located 110~m from the detector, and (ii) HNLs produced in $K$DIF from the BNB target, located 600~m from the detector. Our results are presented in Fig.~\ref{fig:summary}. The red solid line represents the search for $N \to \nu e^+ e^-$ from $K$DAR, while the red dashed line shows the results for the same search but for HNLs produced in $K$DIF from the BNB line. We have used $3 \times 10^{21}$ POT, corresponding to about 5 years of ICARUS data~\cite{Bock:2018nea, Batell:2019nwo, ICARUS_NU_24}. 

For the search for HNLs from $K$DARs, we assume that ICARUS can achieve similar or better performance compared to MicroBooNE in reconstructing the $N \to \nu e^+e^-$ signal. MicroBooNE was able to reduce backgrounds dramatically by exploiting a number of kinematic features of the signal $e^+ e^-$ pair. Most notable is the difference of directions between the detector and (a) production of background-inducing neutrinos in the NuMI target vs.~(b) production of HNLs in the NuMI absorber. Additionally, there is a strong separation between the time of production of the signal and background. These were key to MicroBooNE's success and will be similarly powerful in ICARUS relative to the NuMI beam and absorber. Additional features that are exploited include the opening angle of the $e^+e^-$ pair, as well as the number of hits and the length of the reconstructed objects in the LArTPC~\cite{MicroBooNE:2023icy}. To estimate the HNL efficiencies as a function of mass, we follow the procedure outlined in Ref.~\cite{Kelly:2021xbv}, by recasting a previous MicroBooNE analysis~\cite{MicroBooNE:2020vlq}, assuming that the background rate at ICARUS is the same. In principle, since ICARUS is located $5.5^\circ$ off-axis from the NuMI beamline, the beam-induced neutrino background should be significantly mitigated. Moreover, HNLs produced by two-body decays of kaons at rest will have a fixed energy for a given mass, cf.~Eq.~\eqref{eq:EN}. This will help to discriminate them from neutrino and cosmic-ray-induced background processes. The PMTs surrounding the detector provide $\sim$ns timing resolution, which is a promising
handle to select for HNL signals against
the neutrino background~\cite{Putnam, SBND, SBND:2024vgn}. In this sense, the $K$DAR-ICARUS sensitivities presented here should be considered to be conservative.  

The reconstruction of low-energy electrons may be challenging. We apply two kinematical cuts on top of the HNL efficiencies, namely, the electron energy $E_{e^\pm}>10$ MeV, so that the two tracks can be reconstructed, and the opening angle $\theta_{e^+e^-}>10^\circ$, so that two distinct tracks can be identified. Under these conditions, the ICARUS sensitivity is determined simply by recasting the MicroBooNE search~\cite{MicroBooNE:2020vlq}, as in Ref.\cite{Kelly:2021xbv}, with adjustments for the geometrical area of ICARUS, active volume, and the POT. 

For the search for HNLs produced in $K$DIF in the BNB line, we use the HNL fluxes presented in  Fig.~\ref{fig:flux_HNL} to calculate the expected number of events in the detector. To determine the efficiencies, we performed a Monte Carlo simulation, including energy and angular smearing of electrons and positrons, and applied the same kinematic cuts as in the $K$DAR search to the signal events. Our sensitivity was then derived using a Poisson distribution~\cite{Feldman:1997qc}, corresponding to a 95\% C.L. limit for 2.9 expected signal events and assuming no backgrounds. This is on the aggressive side, but we find that the sensitivity obtained from $K$DAR configuration is still better than that obtained from $K$DIF.

For comparison, we also show in Fig.~\ref{fig:summary} the relevant existing experimental constraints from PIENU~\cite{PIENU:2019usb}, KEK-E89~\cite{Hayano:1982wu}\footnote{An improved limit was reported in Ref.~\cite{Yamazaki:1984sj} from the E104 experiment with higher momentum resolution, more effective background suppression and better particle identification, as compared to the E89 experiment. However, since this is an unpublished result, we only show the official E89 result in Fig.~\ref{fig:summary}. Anyway, the E104 limit is still weaker than MicroBooNE, and therefore, including it will not exclude any additional HNL parameter space.}, PS191~\cite{Bernardi:1987ek}\footnote{According to Refs.~\cite{Arguelles:2021dqn, Gorbunov:2021wua} (see also Refs.~\cite{Kusenko:2004qc, Ruchayskiy:2011aa}), the official PS191 constraint~\cite{Bernardi:1987ek} might have been overestimated. In any case, the entire PS191 curve is inside other exclusion contours and does not affect our discussion.}, T2K~\cite{T2K:2019jwa}, E949~\cite{E949:2014gsn} and MicroBooNE~\cite{MicroBooNE:2023icy} by the shaded regions which are ruled out. There also exist other constraints from individual analyses (not by experimental collaborations), namely, (i) from a fit~\cite{deGouvea:2015euy} to the Michel electron spectrum in muon decay measured by TWIST~\cite{Bayes:2013esa}, (ii) recasting the single electron events in LSND~\cite{Ema:2023buz}, (iii) extrapolating the T2K efficiencies to lower HNL masses~\cite{Arguelles:2021dqn}, and (iv) using the atmospheric neutrino data from Super-K~\cite{Coloma:2019htx}. These additional limits are also shown in Fig.~\ref{fig:summary} by thin solid curves. Future projections from T2K full run~\cite{Arguelles:2021dqn}, DUNE near detector~\cite{Berryman:2019dme}, PIP2-BD~\cite{Ema:2023buz} and SHiNESS~\cite{Soleti:2023hlr} are also shown. Below the pion threshold of 30 MeV, the PIENU limits are expected to be improved by about an order of magnitude by the proposed PIONEER experiment~\cite{PIONEER:2022yag}, not shown here. Also, we do not show the higher HNL mass range from 200 MeV up to the kaon threshold, $m_K-m_\mu\simeq 388$ MeV because the ICARUS sensitivities turn out to be weaker than the existing limits in this range from T2K~\cite{T2K:2019jwa}, E949~\cite{E949:2014gsn}, and  NA62~\cite{NA62:2021bji}. Beyond the kaon threshold, the HNLs can still be produced from heavier $D$ and $D_s$ meson decays; however, the corresponding rates are orders of magnitude smaller than the pion and kaon production rates for the NuMI beam, and more importantly, the $D, D_s$ mesons are too short-lived to reach the NuMI absorber; therefore, we do not consider them here.  

Given the current constraints and the potential of future searches, we will explore the synergy between ICARUS and PIP2-BD. By using the blue star point as a benchmark event in ICARUS, we aim to forecast the number of events PIP2-BD might observe. This joint effort will be crucial in determining whether the HNL hypothesis could account for the signal detected by ICARUS.

It is important to note that part of the allowed parameter space shown in Fig.~\ref{fig:summary}, including the blue star, corresponds to HNLs with lifetimes exceeding 0.1~s. Such HNLs should, in principle, be excluded by BBN constraints~\cite{Sabti:2020yrt, Boyarsky:2020dzc}. However, the BBN constraints are model-dependent, and can be evaded, e.g. by invoking additional HNL decays into dark sector particles before BBN. As long as the HNL decay length is longer than $\sim {\cal O}(100~{\rm m})$, this additional decay channel does not affect our analysis. In general, the laboratory searches serve as an independent test of the HNL hypothesis, irrespective of the cosmological history of HNLs.

\section{Synergy between ICARUS and PIP2-BD}
\label{sec:pip}
In case of a positive detection of HNL at ICARUS, depending on its mass and mixing, other future accelerator neutrino experiments can be used to verify the signal. As shown in Fig.~\ref{fig:summary}, DUNE near detector (with DIF HNLs) is capable of probing the entire ICARUS sensitivity region. On the other hand,  relatively lower HNL masses, e.g. the benchmark point represented by the blue star in Fig.~\ref{fig:summary}, are also accessible at other experiments, such as T2K (full run), PIP2-BD and SHiNESS.  

For illustration, let us compare the ICARUS sensitivity with the PIP2-BD sensitivity derived in Ref.~\cite{Ema:2023buz} using stopped pions and muons. The PIP2-BD detector is placed at $d=18$~m away from the target and the active volume is modeled as a cylinder of 4.5~m height and 4.5~m diameter. 
The mother particles considered are muons and pions. The number of stopped pions (and hence muons) per proton is 0.1. 
For the assumed $1.2\times 10^{23}$ POT, the total number of particles that decay at rest is $N_\pi = N_\mu = 1.2\times 10^{22}$. 

Using the differential decay widths 
\begin{align}
    \begin{split}
        \frac{d\Gamma(\mu \to e \nu_\mu N)}{dE_N} &= \frac{G_F^2 |U_{\mu N}|^2}{12 \pi^3}(3E_N(m_\mu^2 + m_N^2) - 4 m_\mu E_N^2 - 2 m_\mu m_N^2)\sqrt{E_N^2 - m_N^2},\\
        \frac{d\Gamma(\pi \to \mu N)}{dE_N} &= \frac{G_F^2 f_\pi^2 |V_{ud}|^2 |U_{\mu N}|^2}{7 \pi m_\pi^3}((m_\mu^2 +m_N^2)m_\pi^2 - (m_\mu^2 + m_N^2)^2 + 4 m_\mu^2 m_N^2)\\
        &\times \sqrt{m_\pi^4 - 2(m_\mu^2 + m_N^2)m_\pi^2 + (m_\mu^2 - m_N^2)^2}\times \delta\biggl(E_N - \frac{m_\pi^2 + m_N^2 - m_\mu^2}{2 m_\pi}\biggr),
    \end{split}
    \label{eq:prod_channels}
\end{align}
where the muon-induced production is possible for $0 < m_N < m_\mu - m_e$ and the pion-induced production is for $0< m_N< m_\pi - m_\mu$, we can estimate the total number of $e^+e^-$ events from HNL decay as follows~\cite{Ema:2023buz}:~\footnote{These numbers are for Majorana HNL, so a factor of two higher than those quoted in Ref.~\cite{Ema:2023buz} which considered a single (pseudo-)Dirac HNL. }
\begin{align}
    \begin{split}
        N_{ee}^\mu &= 2.8\times 10^5 \left(\frac{|U_{\mu N}|^2}{10^{-6}} \right)^2\left(\frac{m_N}{m_\mu} \right)^6\left(1 - \frac{m_N}{m_\mu} \right)^4 \left(1 + \frac{4 m_N}{m_\mu} + \frac{m_N^2}{m_\mu^2} \right),\\
        N_{ee}^\pi &= 1.6 \times 10^5  \left(\frac{|U_{\mu N}|^2}{10^{-6}} \right)^2\left(\frac{m_N}{m_\mu} \right)^6\frac{m_\pi((m_\mu^2 + m_N^2)m_\pi^2 - (m_\mu^2 + m_N^2)^2 + 4m_\mu^2 m_N^2 )}{m_\mu (m_\pi^2 - m_\mu^2)^2}.\\
    \end{split}
    \label{eq:estimate_PIPII}
\end{align}

For ICARUS, 
we are looking at kaon decays at rest and we have from Ref.~\cite{MicroBooNE:2020vlq} that the number of neutrinos from these decays is 0.085 per POT, considering POT=$0.92\times 10^{20}$, we find $N_K \sim 2.5\times 10^{20}$. The production channel is similar to the pion case in Eq.~\eqref{eq:prod_channels} with the replacement $m_\pi \to m_K$. We can integrate the master formula in Eq.~\eqref{eq:master_formula} with the decay widths given in Eq.~\eqref{eq:prod_channels} to find the total number of $e^+e^-$ events at ICARUS. We get
\begin{equation}
     N_{ee}^K = 553 \left(\frac{|U_{\mu N}|^2}{10^{-6}} \right)^2\left(\frac{m_N}{m_\mu} \right)^6\frac{m_K((m_\mu^2 + m_N^2)m_K^2 - (m_\mu^2 + m_N^2)^2 + 4m_\mu^2 m_N^2 )}{m_\mu (m_K^2 - m_\mu^2)^2}.
     \label{eq:estimate_ICARUS}
\end{equation}
Now we can calculate the ratio of HNL decay events in PIP2-BD and ICARUS. It is a-$|U_{\mu N}|^2 $ independent quantity and hence a function solely of $m_N$. For illustration, we take the point represented by blue star in Fig.~\ref{fig:summary} which is $m_N = 90$~MeV and $|U_{\mu N}|^2 = 3\times 10^{-6}$. For this benchmark point we get
\begin{align}
    \begin{split}
        N_{\text{$K$DIF}}^{\text{ICARUS}} &\sim 9,\\    
        N_{\text{$K$DAR}}^{\text{ICARUS}} &\sim 426,\\
        N_\mu^{\text{PIP2-BD}} &\sim 2300.
    \end{split}
\end{align}

So far, we have only considered the kaon contributions to HNL production for ICARUS. We estimated that including the pion (muon) decay contributions will improve the event numbers and ICARUS sensitivity below 30 ($\sim 60$) MeV by a factor of 5 for $K$DAR searches, which would however be still inside the PIENU (LSND) exclusion region and does not explore new parameter space. Moreover, the mass reconstruction method discussed in Section~\ref{sec:mass_recon} only works for two-body decays, and thus is not applicable for the three-body muon decay. 

\section{Conclusions} \label{sec:con}
HNLs are well-motivated candidates for BSM physics associated with neutrino mass, but there is no unique preference for the HNL mass scale from theory. 
We have shown for the first time that stopped meson sources at accelerator-neutrino experiments have the unique advantage of reconstructing the HNL mass in case of a positive signal in the $N\to e^+e^-\nu$ channel. 
This mass reconstruction is fairly robust against realistic electron energy and angle resolution.
Moreover, even just a handful of events can lead to a reconstruction of the HNL mass at the 5-10\% precision.
Overall, the currently running ICARUS experiment at Fermilab can explore new HNL parameter space in the 70--190 MeV mass range using the stopped kaon source at the NuMI absorber.  
We also explored the synergy between ICARUS and future accelerator-neutrino beam dump facilities like PIP2-BD and DUNE near detector, which could be used to confirm a potential HNL signal at ICARUS.  
\section*{Acknowledgments}
We would like to thank Luighi P. S. Leal and Matheus Martines for useful discussions. G.F.S.A. and P.S.B.D. would like to thank the hospitality of the Fermilab Theory Group, where part of this work was done. We also acknowledge the Center for Theoretical Underground Physics and Related Areas (CETUP*) and the Institute for Underground Science at SURF for hospitality and for providing a stimulating environment, where a part of this work was done. G.F.S.A. is financially supported by Funda\c{c}\~ao de Amparo \`{a} Pesquisa do Estado de S\~ao Paulo (FAPESP) under Contracts No. 2022/10894-8 and No. 2020/08096-0. The work of P.S.B.D. was partly supported by the U.S. Department of Energy under grant No. DE-SC 0017987. The work of K.J.K. was supported by the DOE Grant No. DE-SC0010813. P.A.N.M. is supported by Fermi Research Alliance, LLC under Contract No. DE-AC02-07CH11359 with the U.S. Department of Energy, Office of Science, Office of High Energy Physics. 

\medskip 

{\bf Note Added:} While we were finalizing this work, Ref.~\cite{Chatterjee:2024duf} appeared on arXiv, which also discussed HNL searches at ICARUS. When considering similar production modes (specifically the mass range in which KDIF dominates), we find that their sensitivity curves assuming 15\% efficiency roughly match our sensitivity results. A detailed comparison with the sensitivity results of  Ref.~\cite{Chatterjee:2024duf} is beyond the scope of this work. The main new point of our work is the HNL mass reconstruction using $K$DAR and the synergy with PIP2-BD. 
\bibliographystyle{utcaps_mod}
\bibliography{ref.bib}
\end{document}